# $Sr_4Os_3O_{12}$ — A Layered Osmate(V,VI) that is Magnetic close to Room Temperature

Gohil S. Thakur*[a,b,c], Julia-Maria Hübner,[b] Kati Finzel,[b] Thomas Doert,[b] and Michael Ruck[b,c,d]

*Dedicated to Professor Martin Jansen on the occasion of his 80th birthday.*

**Abstract:** Black crystals of the mixed-valence osmate(V,VI) $Sr_4Os_3O_{12}$ were grown via a gas phase reaction in an evacuated silica tube. The material is attracted to a permanent magnet at room temperature (295±5 K) but loses this property when heated or cooled. In this temperature interval, soft magnetic behavior with vanishing coercivity is observed, but the saturation magnetization is only 0.05 $\mu_B$ per formula unit. Resistivity measurements show that $Sr_4Os_3O_{12}$ is a semiconductor with a small band gap of about 0.26 eV at room temperature. X-ray diffraction on a single-crystal revealed a rhombohedral structure with the space group $R\bar{3}m$ and lattice parameters $a_r$ = 554.64(2) pm and $c_r$ = 2700.1(1) pm at 300(1) K. The compound crystallizes in the $La_4Ti_3O_{12}$ structure type and is isostructural to $Sr_4CoRe_2O_{12}$. The structure is a rhombohedral stack of layered blocks, each of which can be considered as a section of three layers of the cubic perovskite structure cut perpendicular to one of its threefold axes. The Os–O–Os bond angle of about 175° favors superexchange that leads to antiferromagnetic coupling between the osmium atoms with $5d^3$ and $5d^2$ configurations. $Sr_4Os_3O_{12}$ exhibits strong lattice dynamics. One aspect is the antiferro-rotative lattice modes of the corner-sharing [$OsO_6$] octahedra, which freeze upon cooling. At 100 K, an ordered, yet twinned triclinic superstructure is formed. Strong spin-orbit coupling and other effects cause deformations of the [$OsO_6$] octahedra that differ significantly at 300 K and 100 K, suggesting a change in the electronic configuration. Competing ground states could also explain the temperature-dependent band gap and the magnetic fluctuations manifested in the magnetization peak at room temperature.

## Introduction

Emergence of quantum phenomena in systems containing the platinum group metals has accelerated research on ruthenates and iridates based on perovskite motifs. Ruddlesden-Popper phases $Sr_{n+1}M_nO_{3n+1}$ ($M$ = Ru, Ir; $n$ = 1, 2, 3, ...) were found to possess intriguing physical properties.[1–7] Unfortunately, a similar series does not exist for osmates. The only known homologous structural series of osmates is $Sr_{2n-1}Os_nO_{5n-2}$ for which only the members $n$ = 3 and 5 are known so far.[8,9] Nonetheless, osmates can also exhibit exotic electronic and magnetic states due to several competing parameters such as strong spin-orbit coupling (SOC), Coulomb repulsion, crystal field, Hund's coupling, and Jahn-Teller distortion.[10–14] Unfortunately, osmates are among the least explored oxo-compounds of the *d*-block elements. Ternary and quaternary osmates are difficult to synthesize because sensitivity to air and moisture can lead to decomposition, some of them are only metastable, and the intermediate or by-product $OsO_4$ is toxic. Nonetheless, there are several stable osmates with interesting physical and chemical properties, especially from the past two decades. Illustrative examples are superconductivity in the $AOs_2O_6$ series ($A$ = K, Rb, Cs),[15–17] a metal-insulator transition in $Cd_2Os_2O_7$,[18] a Slater transition in $NaOsO_3$,[19] a ferroelectric state in $LiOsO_3$,[20] strong Jahn-Teller compression and spin-pairing in the $5d^2$ systems $Na_9Bi_5Os_3O_{23}$[21] and $Na_2OsO_4$,[22] as well as high-temperature magnetic ordering in $Sr_3OsO_6$[23], $Sr_3CaOs_2O_9$[24] $Sr_8Os_{6.4}O_{24}$[25] and $NaOsO_3$.[26] In particular, the Sr-Os-O system has allowed access to several new compounds, many of which are based on the perovskite or pyrochlore lattice, e.g. $Sr_3OsO_6$,[23] $Sr_8Os_{6.3}O_{24}$,[25] $Sr_5Os_3O_{13}$,[9] $Sr_7Os_4O_{19}$,[27] $Sr_2OsO_5$,[27] $Sr_9Os_5O_{23}$,[8] $Sr_{11}Os_4O_{24}$,[28] $SrOsO_4$,[29] $Sr_2Os_2O_{6.4}$,[30] or $Sr_3Os_4O_{14}$.[31] Among them, the perovskite-type based osmates have emerged as magnetic materials such as ferro(ferri-)magnetic insulators or antiferromagnets with surprisingly high Curie or Néel temperatures.[23–26] Fully-compensated ferrimagnetic materials with half-metallic character are of special interest as they combine the advantage of a small (zero) magnetic moment of a antiferromagnet with the ease of realizing magnetic reading of a ferromagnet and are not affected by any stray field.[32–35]

Here we report the discovery of the new mixed-valence osmate(V,VI) $Sr_4Os_3O_{12}$, which crystallizes as a layered defect variant of the perovskite type. Remarkably, this material undergoes at least one magnetic transition close to room temperature (295 ±5 K), which generates a significant macroscopic magnetization exclusively in this narrow temperature range.

## Results and Discussion

### Synthesis

Black hexagonal block-shaped crystals of the new osmate(V,VI) $Sr_4Os_3O_{12}$ were obtained by a reaction of strontium oxide (SrO) and strontium peroxide ($SrO_2$) with osmium dioxide ($OsO_2$) in a silica ampoule at 800 °C. $OsO_2$ and $SrO/SrO_2$ had to be kept separately in open crucibles stacked one on top of the other inside the ampoule. This suggests that the reaction proceeds through gaseous $OsO_4$, which enables transport of osmium to the strontium oxide at reaction temperature. A similar procedure has been earlier adopted to successfully synthesize single-crystals of $Sr_8Os_{6.3}O_{24}$[25] and $Sr_5Os_3O_{13}$[9]. In contrast, a direct reaction of an intimate stoichiometric mixture of the starting materials lead to a mixture of phases with $KSbO_3$-type, $Sr_{1-x}OsO_3$ being predominant impurity. Adding a small amount of $Na_2O_2$ (5 at.%, kept in SrO containing crucible) greatly improved the yield, though the by-phases could not be entirely eliminated. It is emphasized that no trace of sodium was detected by EDX spectroscopy in any of the $Sr_4Os_3O_{12}$ crystals, while the molar ratio Sr : Os of 4 : 3 was confirmed (Figure S1 of the Supporting Information). $Na_2O_2$ increases the oxygen partial pressure, which promotes the formation of osmium tetraoxide ($OsO_4$) and creates an oxidizing atmosphere that stabilizes higher oxidation states of osmium. Absence of $Na_2O_2$ leads to the formation of a mixture of previously reported $Sr_4Os_{3.2}O_{12}$ (= $Sr_8Os_{6.4}O_{24}$) and unreported $Sr_{1-x}OsO_3$ ($0.7 \leq x \leq 0.8$), in which the average oxidation state of osmium is lower than in the title compound (Table S1 in SI). The described synthesis method is only fairly reproducible. $Sr_4Os_3O_{12}$ formed also in a salt flux growth by reacting $SrO_2$, Os metal and anhydrous $SrCl_2$ (3 : 1 : 4 molar ratio) in an evacuated ampoule at 800 °C. However, only a few crystals were obtained after the flux was dissolved in water aided by sonication.

[a] Prof. Dr. G. S. Thakur
Department of Chemical Sciences, Indian Institute of Science Education and Research, Berhampur, 760010, India
E-mail: gsthakur@iiserbpr.ac.in
[b] Dr. G. S. Thakur, Dr. J. Hübner, Dr. K. Finzel, Prof. Dr. T. Doert, Prof. Dr. M. Ruck
Faculty of Chemistry and Food Chemistry
Technische Universität Dresden, 01062 Dresden, Germany
[c] Prof. Dr. G. S. Thakur, Prof. Dr. M. Ruck
Würzburg–Dresden Cluster of Excellence 'ct.qmat',
Technische Universität Dresden, 01062 Dresden, Germany
[d] Prof. Dr. M. Ruck
Max Planck Institute for Chemical Physics of Solids
Nöthnitzer Straße 40, 01187 Dresden, Germany

Reproducibility issues in the synthesis of ternary osmates have several reasons. There are obviously many compounds in the AE-Os-O systems (AE = Sr, Ba), and most of the thermodynamically stable ternary phases form between 700 and 1000 °C.[8,9,25,27,36–39] The volatility of $OsO_4$ can reduce the amount of osmium available for the reaction in the mixture of starting materials. Partial pressures of oxygen and gaseous $OsO_4$ as well as the total amount of starting materials strongly influence composition of the product(s). For example, $Sr_2OsO_5$[27] readily forms in high yield via a sealed tube reaction of stoichiometric amounts of $OsO_2$, $SrO_2$ and SrO at 800 °C when the total quantity is small (~ 200 mg). However, when synthesis of a larger batch is attempted a new phase '$Sr_7Os_3O_{15}$' forms under same reaction conditions. The situation is similar for the crystal growth of $Sr_9Os_5O_{23}$ (average oxidation state of osmium +5.6),[8] the larger the relative volume of the ampoule, the more $Sr_5Os_3O_{13}$ (average oxidation state +5.33) is formed. In the present case, the total amount of raw material was restricted to a maximum of ~ 200 mg to avoid excessive vapor pressure.

*Magnetic Properties*

Figure 1a shows the temperature dependence of the magnetic moment of $Sr_4Os_3O_{12}$ measured on some loose crystals in magnetic fields up to $\mu_0H$ = 7 T. The compound is only weakly ferromagnetic-like in the narrow range between 280 and 320 K, where a peak in the magnetic moment is observed in all applied fields (maximum at 290±5 K). Consistent with the magnetic data, the crystals attract towards a permanent iron neodymium magnet at room temperature (295 K, 23 °C) but loose this property when heated slightly above room temperature indicating a paramagnetic state.

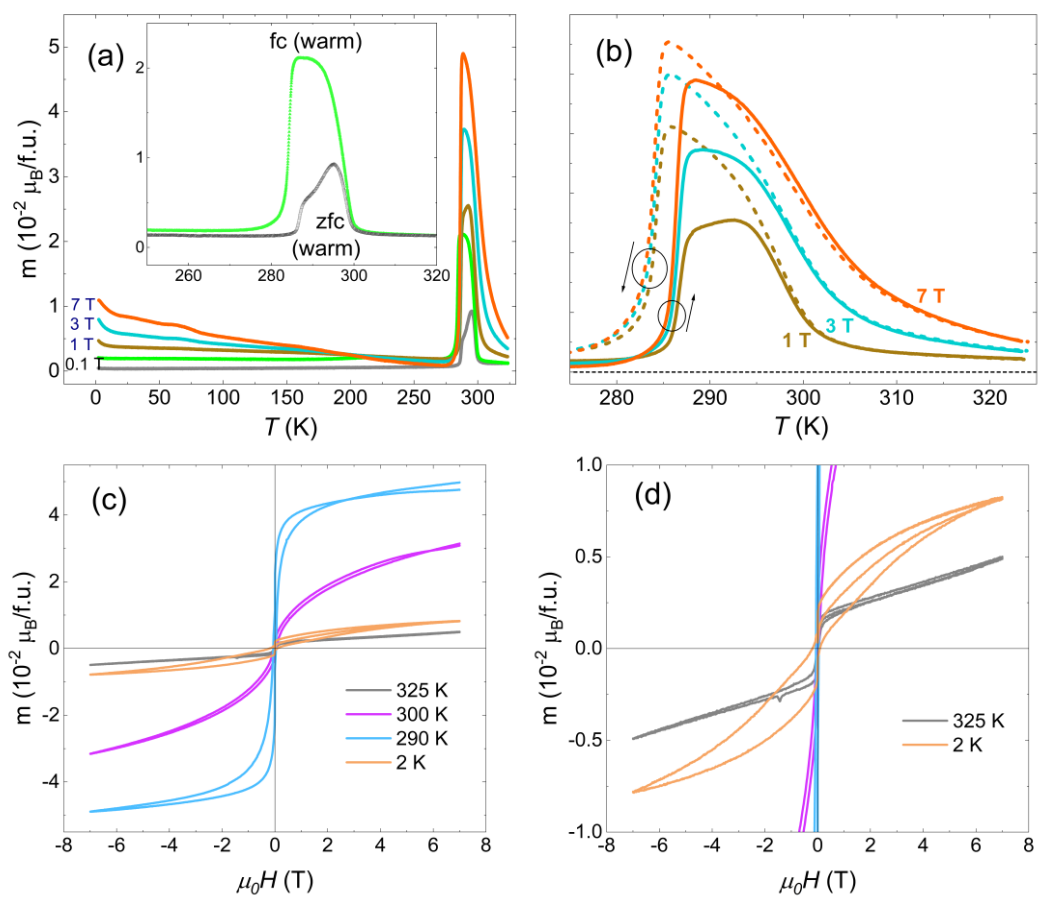

**Figure 1**. (a) Temperature dependence of the magnetic moment of $Sr_4Os_3O_{12}$ crystals in several external magnetic fields. The inset shows the zero-field-cooled (zfc) and field-cooled (fc) curves near the magnetic transition. (b) Magnification of the magnetic moment near the magnetic transition during warming and cooling. (c) Field-dependence of the magnetic moment at different temperatures. (d) Magnification of the hysteresis loops at 2 K and 325 K.

There is a marked hysteresis in warming and cooling runs in all applied fields (Figure 1b), namely that the magnetization maximum lies a few degrees lower in cooling than in warming runs. This indicates that the transition is of first order in nature. Moreover, the magnetic transition broadens with increasing the applied field. However, the temperature of the steep edge is rather consistent among the cooling and warming cycle and does not change with increasing field. The field dependence of the magnetic moment reproduces this behavior (Figure 1c,d). The magnetic moment at 325 K is nearly linear with negligible coercivity, but increases strongly below 300 K and reaches its maximum at 290 K, before dropping to low values again at lower temperatures. The magnetic moment at 290 K shows saturation in moderate fields but with a small saturation moment of only 0.05 $\mu_B$ per formula unit (f.u.). The magnetization curves at 2 K, 300 K, and 325 K are also non-linear, but do not show a clear tendency to saturation. None of the curves indicates a remnant magnetization of the sample at zero field. The shape of the magnetization curves with the sharp drop at the low-temperature edge of the peak resembles an antiferromagnetic ordering emerging from a paramagnetic high-temperature phase. However, the maximum of the magnetic moment is broadened in all fields (Figure 1b) and even shows a shoulder in the warming cycle after zero-field cooling (zfc; inset of Figure 1a), which is unusual for a conventional antiferromagnetic transition. The observed attraction towards a permanent magnet indicates a small intrinsic ferromagnetic component. Similar features in magnetization were observed for (half-)metallic ferrimagnets such as $Mn_2PtAl$[42], $Mn_2PtGa$[43], (CrFe)S[35], or $Ni_{50}Mn_{34}In_{16}$,[44] where a jump in magnetization followed by a sudden drop to near zero moment is reported, yet over temperature ranges of tens (or hundreds) of kelvin. However, in the semiconductor $Sr_4Os_3O_{12}$, the exact nature of the magnetic transition cannot be clearly assigned based on limited data. Additional detailed structural and magnetic studies like neutron diffraction, AC susceptibility and X-ray magnetic circular dichroism are required.

Trigonal lattices of magnetic atoms with antiferromagnetic exchange tend to show magnetic frustration. In $Sr_4Os_3O_{12}$, however, the osmium atoms of one layer are separated by 555 pm (i.e. lattice parameter *a*) and their coordination polyhedra have no direct connection (see later sections). Across the gap between the layer blocks, the osmium atoms are facing each other with a distance of 434 pm, which allows for a through space interaction and thus potentially for antiferromagnetic coupling between adjacent layer blocks. With regard to the interatomic distances, however, both interactions are considered too weak to be responsible for the magnetic order above room temperature.

A complementary X-band electron spin resonance (ESR) experiment showed no distinct signal in the temperature range from 200 to 360 K. A change of the resonator frequency was found around 290 K which may arise from a change of the magnetization of the sample.

*Electrical Conductivity*

The electrical resistance in dependence of temperature was measured on a small $Sr_4Os_3O_{12}$ crystal using a four-point probe setup (Figure 2 and S2). $Sr_4Os_3O_{12}$ is a semiconductor down to at least 60 K with a small band gap of $E_g$ = 0.26 eV in the temperature range from 170 to 320 K. This is a typical value for highly oxidized osmates such as $Ba_2NaOs^{VII}O_6$ (~0.30 eV)[45] or $Na_9Bi_5Os^{VI}_3O_{24}$ (0.177 eV).[21] The resistivity at 300 K ($\rho_{300\,K}$) amounts to $5.9 \cdot 10^{-2}$ Ωm. The Arrhenius type plot (Figure 2, inset) reveals a non-linear behavior of the resistance with an additional, distinct temperature regime between 60 and 85 K. In the low-temperature region, $E_A$ decreases to 0.02 eV. The cross-over point between the two regimes occurs at approximately 125 K. Similar changes in activation energy were reported for osmium dipnictides at high temperatures (measurement range: room temperature to 700 K).[46] This non-linear behavior could hint to changes in the crystal field or an increasing charge carrier mobility.

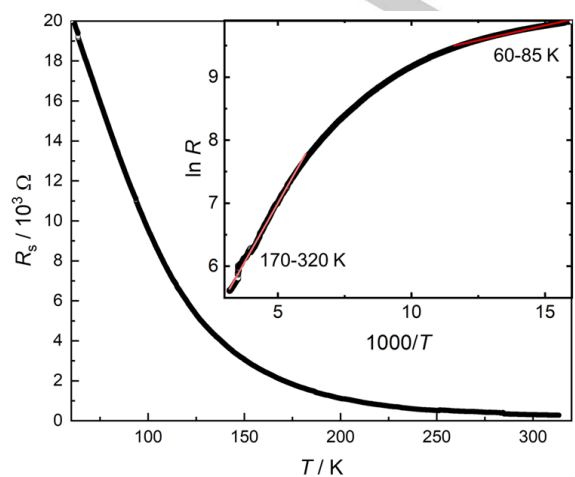

**Figure 2.** Temperature dependent electrical resistance of a $Sr_4Os_3O_{12}$ crystal measured using a four-point probe and corresponding Arrhenius type plot (inset).

*Crystal Structure*

The powder X-ray diffraction pattern of $Sr_4Os_3O_{12}$ (PXRD, Figure 3) looks very similar to that of $Sr_4CoRe_2O_{12}$[41] and shows all the reflections pertaining to that phase. Additionally, very weak reflections of an unknown by-phase are observed (marked with an asterisk). A Lebail fit of the PXRD pattern yielded a rhombohedral lattice with $a \approx 556$ pm and $c \approx 2711$ pm. A detailed inspection of the diffractogram revealed splitting of reflections at high diffraction angles (see below).

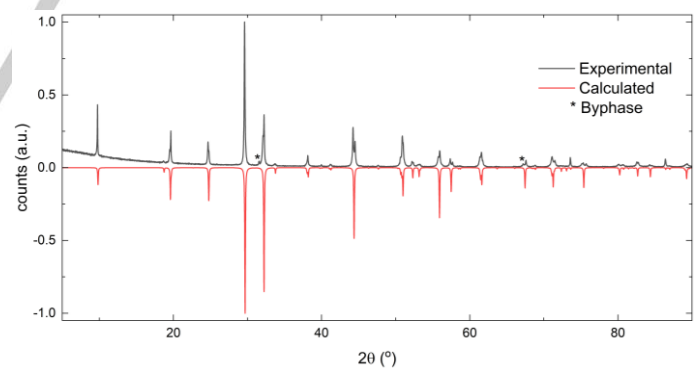

**Figure 3.** Room temperature powder diffraction pattern of $Sr_4Os_3O_{12}$ alongside the powder pattern calculated using single crystal data.

The indexing was confirmed by X-ray diffraction on a single-crystal at 300(1) K (Table S2). The primary structure solution indicated that $Sr_4Os_3O_{12}$ adopts the space group $R\bar{3}m$ with lattice parameters $a_r$ = 554.64(2) pm and $c_r$ = 2700.1(1) pm and a unit cell volume of $V_r$ = 719.34(6)·10$^6$ pm$^3$. The corresponding atomic parameters are given in Table S3. Based on this structure model, the compound crystallizes in the $La_4Ti_3O_{12}$ structure type and is isostructural to $Sr_4CoRe_2O_{12}$. These structures can be generally understood as *B*-site deficient hexagonal perovskites with general formula $A_4\square B'B_2O_{12}$, where *B* and *B'* sites can be occupied by either one or more type of transition metals. Such phases (e.g. $Sr_4Re^{VII}_2MO_{12}$ with $M$ = Mg, Co, Ni, Zn) were first identified by *Ward* et al. in 1961.[47] The $Sr_4Os_3O_{12}$ structure is a rhombohedral stack of layered blocks (Figure 4a,b), each of which can be considered as a section of three layers of the cubic perovskite structure cut perpendicular to one of its threefold axes.

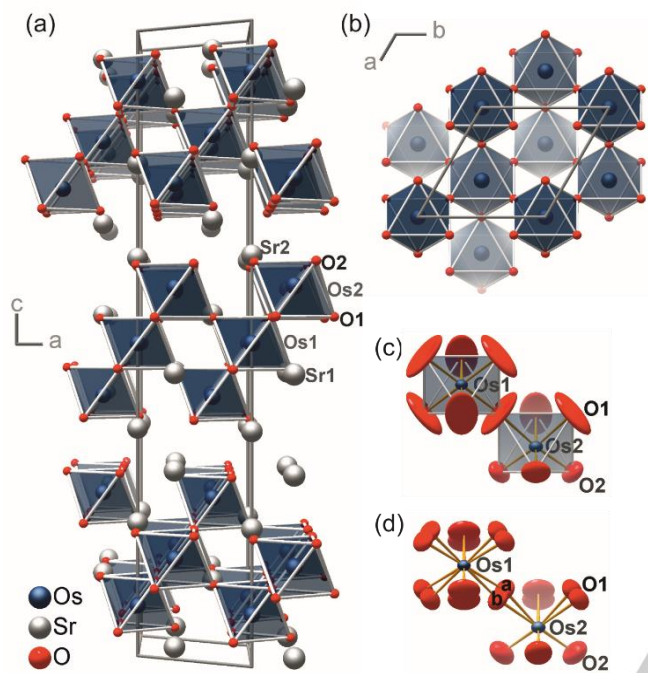

**Figure 4.** (a) Layered crystal structure of $Sr_4Os_3O_{12}$ in $R\bar{3}m$ at 300 K. (b) Top view of a block of three layers of corner-sharing [$OsO_6$] octahedra (Sr atoms omitted). (c) Coordination polyhedra of the osmium atoms shown with displacement ellipsoid (99% probability). (d) Same with split positions for O1.

The oxygen and the strontium atoms together form a dense sphere packing with the stacking sequence $(hcch)_3$ (Jagodzinski nomenclature).[48] Osmium atoms fill a quarter of the octahedral voids in the blocks, specifically those that are defined exclusively by oxygen atoms. The layer of octahedral voids between two blocks remains empty. The coordination polyhedra of Sr1 and Sr2 are a cuboctahedron and an anticuboctahedron, respectively. The octahedron around the Os1 atom (Wyckoff position 3$b$, site symmetry $\bar{3}m$) in the middle layer of a block is almost regular with Os1–O1 bond lengths of 194 pm and O1–Os1–O1 bond angles of 88°, 92° and 180° (Table S4). The octahedron edges perpendicular to the $c$-axis are 10 pm longer than the others, indicating a trigonal compression. The Os2 atom (Wyckoff position 6$c$, site symmetry 3$m$) in the upper and lower layers of a block is located outside the center of its coordination polyhedron. The distances Os2–O1 and Os2–O2 are 202 pm and 184 pm, in accordance with the different bond strength towards bridging and terminal ligands. The bond angles are O1–Os2–O1 86°, O2–Os2–O2 96°, O1–Os2–O2 89° and 173°. The octahedron around Os2 is also slightly compressed along the trigonal axis.

The average oxidation state of $+5^1/_3$ for osmium could be resolved as $Os^{VI}$ on 3$b$ and $Os^V$ on 6$c$. However, the calculated bond valence sums[49] of 5.3 for Os1 and 5.53 for Os2 do not support a charge ordering. In literature, an intervalent state was also reported for $Sr_9Os_5O_{23}$,[8] while mixed valence with a clear charge distribution was found for $Sr_5Os_3O_{15}$.[9] Under these premises, $Sr_4Os_3O_{12}$ should be a metal, which is not the case (see above). Moreover, the displacement ellipsoid of O2, i.e. the atom connecting Os1 and Os2 atoms (angle Os1–O1–Os2 175.8°), is an exceptionally large, flat disk perpendicular to the Os–O bonds (Figure 4c), which indicates a serious problem with the rhombohedral structure model (despite $R_1$ = 0.024 and $wR_2$ = 0.049 for all data). Taking into account that a Jahn-Teller distortion can be expected for osmium atoms with 5$d^3$ and 5$d^2$ configurations (see below), a lower symmetric structure appears possible. Twin refinement in the monoclinic space group $R\,2/m\,1$ (alternative setting of $C2/m$) of the triclinic space group $R\bar{1}$ using the same unit cell, however, yielded no improvement of the model. Therefore, single-crystal X-ray diffraction data were also measured at 100(1) K (Table S2). Inspection of the diffraction pattern revealed splitting of reflections at high diffractions angles and superstructure reflections that indicate an enlarged unit cell (Figure 5).

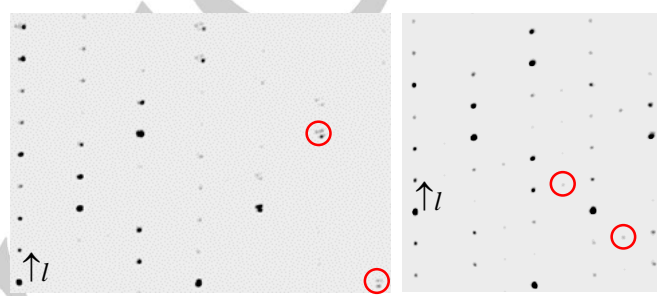

**Figure 5.** Sections of the diffraction pattern of a $Sr_4Os_3O_{12}$ crystal measured at 100 K. Left: $h2l$ layer (rhombohedral indexing) showing splitting of reflections at high diffraction angles. Right: $2kl$ layer showing superstructure reflections.

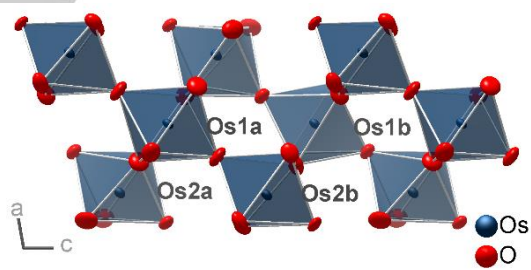

**Figure 6.** Layer package of corner-sharing [$OsO_6$] octahedra in the crystal structure of $Sr_4Os_3O_{12}$ in $C\bar{1}$ at 100 K. The Sr atoms were omitted. Ellipsoid for 99% probability.

Refinements of the 100 K data were performed in the triclinic space group $C\bar{1}$ with lattice parameters $a_t$ = 1816.80(6) pm, $b_t$ = 556.04(1) pm, $c_t$ = 961.99(3) pm, $\alpha_t$ = 89.833(3)°, $\beta_t$ = 99.944(3)°, $\gamma_t$ = 90.279(2)°, $V_t$ = 957.21(4)·$10^6$ pm$^3$. The metrical change upon cooling from 300 K to 100 K is about +1 pm in the base of the layer and –5 pm for the thickness of one-layer package. The choice of the $C$-centered instead of a primitive cell allows a better comparison with the rhombohedral structure, since the $C$-centered base corresponds to the orthohexagonal setting. The transformation of axes from the rhombohedral cell is $a_t = (-2a_r - b_r + 2c_r)/3$, $b_t = -b_r$ and $c_t = 2a_r + b_r$. Twin law according to a threefold along $c_r$ was applied, but without constraints or restraints. The result was an ordered structure model with reasonable displacement parameters (Figure 6 and Table S5). The most striking difference to the room temperature model is the antiferro-rotative tilting of the corner-sharing octahedra in the perovskite-type layers. The Os1–O1–Os2 angles are reduced down to 156°.

The interatomic distance from Os1a and Os1b (both on $\bar{1}$) to the surrounding oxygen atoms are 194 to 197 pm and thus surprisingly homogeneous (Table S6). The O–Os1a/b–O angles deviate by about 2° from 90° and thus the octahedron is almost regular. The distances of Os2a and Os2b to the terminal oxygen atoms range from 181 to 186 pm, those to the bridging oxygen atoms are 202 to 206 pm. The bond angles have relative wide ranges O1–Os2–O1 82–90°, O2–Os2–O2 93–101°, O1–Os2–O2 86–91° and 169–179°. Thus, the polyhedra around Os2a and Os2b are more distorted than at room temperature.

It is remarkable that the Os–O distances appear to be longer at 100 K than at 300 K, which could be an artifact of the rhombohedral structure model. Actually, a Fourier analysis of the 300 K data reveals two distinct maxima of the electron density close to the O1 position. Refinements in the same rhombohedral space group (Table S1 and S6), but with split positions for O1a and O1b (separated by 45 pm), resulted in displacement ellipsoids of the O1 atoms are now similar to that of O2 (Figure 4d). An ordered structure model was constructed, with the constraint that the O–Os–O bond angles must not deviate too much from 90° (Figure 7). This also shows an antiferro-rotative distortion with Os1–O1–Os2 angles of 166° and 164°. The octahedron around Os1 is now axially elongated with distances Os1–O1a 197 pm (2×) and Os1–O1b 193 pm (4×). The effect at Os2 is similar with Os2–O1a 202 pm and Os2–O1b 204 pm.

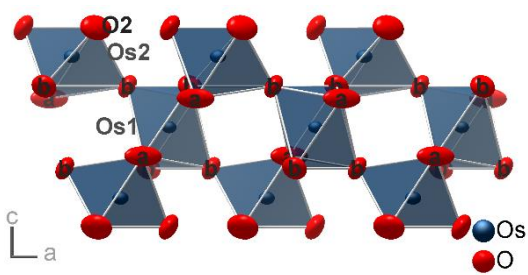

**Figure 7.** Ordering model for a layer package of corner-sharing [OsO$_6$] octahedra in the crystal structure of Sr$_4$Os$_3$O$_{12}$ in $R\bar{3}m$ (T = 300 K) based on split positions O1a,b. The Sr atoms were omitted. Ellipsoid for 99% probability.

Overall, Sr$_4$Os$_3$O$_{12}$ exhibits extensive lattice dynamics. One aspect of this are librations of the octahedra, which are cooperative in antiferro-rotative lattice modes. There also appear to be deformations of the octahedra, probably of electronic origin. At 300 K, the rhombohedral structure model (with split positions) seems to be an adequate description for the crystal structure in the temporal and spatial averages. At 100 K, at least the librational part of the dynamics is frozen and an ordered superstructure is formed.

*Electronic Structure*

As has been found in several previous theoretical studies, DFT-based methods usually fail to reproduce the experimentally found small band gap for highly oxidized osmium compounds.[41,50] This is also the case here. In the electronic band structure (including SOC) based on the initial rhombohedral model at 300 K and the refined model at 100 K, bands cross the Fermi energy $E_F$. The calculated density of states (DOS, Figure 8) has even a local maximum at $E_F$. In previous calculations, high values for the Coulomb repulsion $U$ were used to model a gap and then interpreted as strong electronic correlations.[45,51–53] But also more refined explanations, such as spin-orbital Jahn-Teller bipolarons,[50] have recently been reported in the double perovskite system Ba$_2$Na$_{1-x}$Ca$_x$OsO$_6$ (0 < x < 1).

The successively integrated DOS values show that the general occupation scheme is quite similar as both integrals exhibit similar behavior and integrate to 80 electrons, which means that the valence regions comprises 4 × 2 e$^-$ from the Sr 5s orbitals, 3 × 8 e$^-$ from the Os 6s and 5d orbitals, and 12 × 4 e$^-$ from O 2p orbitals.

The O 2s states are located at lower energy values. For both structural models, the atomic partial charges are about Sr$^{+0.7}$, Os$^{+1.5}$ and O$^{-0.6}$, i.e. much lower than the oxidation states. However, these are only a crude estimate, since no real space analysis has been performed. The main difference between the DOS of the two structure models of Sr$_4$Os$_3$O$_{12}$ lies in the dispersion. The 100 K model exhibits a more homogeneous and less energetically extended valence region than the 300 K model. This hints at more localized states. Figure 9 shows the $l$ = 2 orbital contributions for the Os atoms in both structure models. In the 100 K structure model the Os 5d contributions are more balanced compared to the spread in the 300 K structure. These small differences in the electronic structure are associated with the chemical bonding between Os and O and might cause the unusual temperature dependent magnetic behavior of the compound.

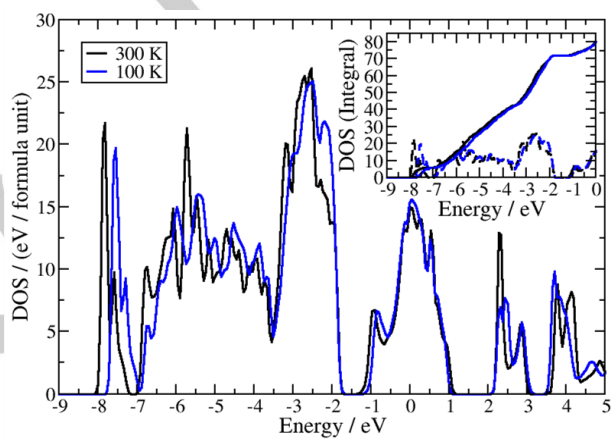

Figure 8. Fully relativistic electronic density of states per formula unit (DOS) for Sr$_4$Os$_3$O$_{12}$ for the two different structural models at 300 K (shown in black) and 100 K (shown in blue). The inset shows the successively integrated DOS for the occupied states in the valence region.

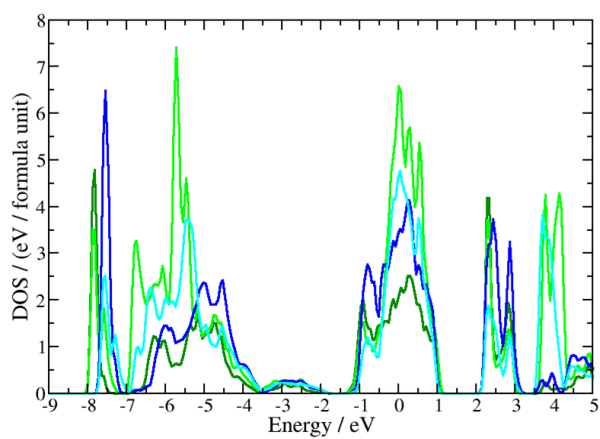

Figure 9. Partial density of states per formula unit (p-DOS) for Sr$_4$Os$_3$O$_{12}$ for the Os $l$ = 2 states of the structure models at 300 K (green and light green) and 100 K (blue and light blue).

Additional qualitative considerations can also be made with regard to the chemical bonding in Sr$_4$Os$_3$O$_{12}$. Osmium atoms in oxidation states +V (Os2) and +VI (Os1) are 5d$^3$ and 5d$^2$ systems, in an octahedral crystal field with $t_{2g}^3$ and $t_{2g}^2$ configurations. A crystal field splitting $t_{2g}$-$e_g$ of about 4 eV can be expected, based on the situation in Ba$_2$CaOsO$_6$ (4.4 eV, Os$^{VI}$–O 192 pm).[51,54] In addition, the electrons of the heavy atom osmium (atomic number 76) experience strong spin-orbit coupling (SOC). SOC leads to a splitting of the degenerate $t_{2g}$ orbitals into a stabilized (by $-0.5\ \lambda$) quartet with total angular moment $J_{eff} = 3/2$ and a destabilized (by $+\lambda$) $J_{eff} = 1/2$ doublet. On the other hand, distortions of the octahedral coordination, either caused by the Jahn-Teller effect or by other chemical reasons (e.g. layer termination), also split the $t_{2g}$ orbitals. A trigonal compression yields a stabilized $a_{1g}$ and two destabilized $e'$ orbitals, while a tetragonal elongation has the opposite effect. In addition, SOC can have an influence on Jahn-Teller distortions by either generating or suppressing them.[10,50,55] Because of the corner-sharing of the [$OsO_6$] octahedra in the structure of $Sr_4Os_3O_{12}$, the Jahn-Teller distortions can influence each other. Moreover, the entire system could be dynamic. Electron hopping interchanges the oxidation states on the osmium atoms and results in small variations of the interatomic distances and a dynamic Jahn-Teller effect. Such excitations in which electron hopping is coupled to vibrational modes of the lattice are known as polarons.[50] In fact, the structure investigations hint at a strong lattice dynamics of $Sr_4Os_3O_{12}$.

At the current state of research, various scenarios for the spin distribution in $Sr_4Os_3O_{12}$ can be outlined, but all of them are speculative. In the literature, even complex scenarios for the explanation of unusual magnetic behavior of osmates(VI) are discussed, e.g. a ferro-octupolar order in $d^2$ double perovskites $Ba_2MOsO_6$ ($M$ = Ca, Mg, Zn)[12] or anti-Hund low-spin states.[13,21] The significant difference in the deformation of the [$OsO_6$] octahedra at 300 and 100 K could indicate a change in the electronic configuration. Competing ground states and a spin crossover would also explain the temperature-dependent band gap and the magnetic fluctuations manifested in the magnetization peak at room temperature. Further investigations including neutron diffraction, high-temperature magnetization measurements, Raman and X-ray absorption spectroscopy as well as band structure calculations on a high level of theory are needed for an understanding of this intriguing compound.

## Conclusions

We have synthesized the new ternary osmate $Sr_4Os_3O_{12}$, which crystallizes in a layered hexagonal perovskite structure. The compound is essentially weakly ferromagnetic-like in the narrow range between 280 and 320 K, where a peak in the magnetic moment is observed in all applied fields up to 7 T (maximum at 290±5 K), and is presumably antiferromagnetic below 280 K. Understanding this unusual magnetic phenomenon and, related to it, the actual electronic situation requires a series of additional experiments and theoretical studies. $Sr_4Os_3O_{12}$ could prove to be a platform compound to study the complex interplay of strong spin-orbit coupling and electronic correlation in combination with valence fluctuation and lattice dynamics.

## Experimental Section

*Attention! Osmium tetroxide is highly toxic. Exposure to $OsO_4$ can damage the skin, eyes and respiratory organs. The toxicity of strontium osmates is not yet known.*

**Synthesis.** Coarse crystalline sample of the composition $Sr_4Os_3O_{12}$ was obtained by heating a mixture of stoichiometric amount of $SrO_2$ and $SrO$ kept in one alumina crucible and $OsO_2$ in a separate crucible. The crucibles were placed one over the other inside a fused silica tube which was subsequently sealed under vacuum. Adding a trace amount of $Na_2O_2$ in the crucible containing $SrO_2$, resulted in a greater yield of the desired product. The tubes were heated in a slanted tube furnace at 800 °C for 48 hours. The heating and cooling rate were maintained at 100 K/h. The weighing and mixing of the starting mixtures were done in an Ar filled glove box (MBraun; $p(O_2)/p^0$ < 1 ppm, $p(H_2O)/p^0$ < 1 ppm). Black reflective single crystals with a hexagonal block like morphology were obtained along with some polycrystalline matrix. Crystals were separated from the product mixture by ultra-sonication in methanol for few minutes. Crystals with clean surface were subjected to single-crystal and powder X-ray diffraction, electron microscopy and magnetic characterization.

**Composition analysis.** The elemental composition of the crystals was determined using a scanning electron microscopy (SEM) equipped with an energy dispersive X-ray (EDX) analyser (SU8020, Hitachi; Silicon Drift Detector X-Max$^N$, Oxford). Several different crystals from different batches were analyzed at five different points each. All the crystals showed an average metal ratio Sr:Os of 4:3 (Figure S1).

**Structure determination:**

**Structure determination:** Crystal structure determination was carried out using single crystal X-ray diffraction (SCXRD). Suitable single-crystals were selected under an optical microscope and mounted in polymer loops. Data sets were collected on a Rigaku Synergy S diffractometer with Ag-K$\alpha$ radiation ($\lambda$ = 56.087 pm) and a hybrid pixel array detector (Dectris Eiger2 1M) at different temperatures using the CrysAlis software package[56]. Structure solution and refinements have been performed with JANA2020[57] and ShelXL.[58,59] Diamond 4[60] was used for visualization. Crystallographic data have been deposited with Fachinformationszentrum Karlsruhe, D-76344 Eggenstein-Leopoldshafen (Germany) and can be obtained on quoting the depository numbers CSD-3000544 (300 K) and CSD-3000545 (100 K).

Powder X-ray diffraction (PXRD) data were collected using an X'Pert Pro diffractometer (PANalytical, Bragg-Brentano geometry, curved Ge-(111) monochromator, fixed divergence slits, PIXcel detector) using CuK$\alpha_1$ radiation ($\lambda$ = 154.059 pm). Several randomly picked crystals were ground to powder and fixed on a single-crystal silicon sample holder using a minimal amount of Apiezon-H grease. The data was collected at 296(1) K in the range $3° \leq 2\theta \leq 90°$.

**Magnetic measurements:** Magnetic data on a collection of randomly oriented loose single crystals were recorded on a vibrating sample magnetometer in a 9T Cryogen Free Measurement System, CFMS (Cryogenic Ltd.). The data were collected during cooling/warming cycle in Zero field cooled (ZFC) and field cooling (FC) protocol under a range of applied field $\mu_0 H$ = 0.1 and 7 T and in the temperature range of 2–325 K. Isothermal magnetization was measured at $T$ = 2, 290 and 300 and 325 K in an applied magnetic field of ± 7 T. No diamagnetic corrections were made.

**Electrical resistance:** The sample (0.7 mm x 0.6 mm x 0.3 mm, Figure S2) was fixed to the puck using varnish (GVL Cryoengineering, IMI 7031) and contacted with platinum wires (0.025 mm diameter, GoodFellow Cambridge Limited) using silver epoxy (Plano GmbH). The electrical resistance was measured (DC resistivity probe) in a temperature range from 60 to 320 K by using the four-point probe method in a Cryogenic Free Magnet System (Cryogenics). Below 60 K, the resistance was too high to be measured with this setup.

**Electronic band structure calculations:** Full-relativistic DFT-based calculations of the electronic band structure of $Sr_4Os_3O_{12}$ for the 100 K and the 300 K structural model were performed using a modified version of the FHI-aims package .[61] The modification consists in an implementation of the bifunctional formalism for the exchange energy (FB16).[62,63] In contrast to density functional theory where the electron density is the only variable,[64–66] the bifunctional formalism[62,63] employs two separate variables, being the electron density and the respective formal density functional derivative, the potential. The DOS was calculated on a 11 × 11 × 2 (300 K) and a 2 × 6 × 4 (100 K) k-points grid in the Brillouin zone.

## Acknowledgements

The authors thank Dr. Jörg Sichelschmidt (MPI CPfS Dresden) for the ESR measurement. We acknowledge financial support by the Deutsche Forschungsgemeinschaft through the Collaborative Research Center SFB 1143 (Project No. 247310070) and the Würzburg-Dresden Cluster of Excellence *ct.qmat* (EXC 2147, Project No. 39085490).

# ARTICLE

**Entry for the Table of Contents**

# FULL PAPER

A new highly oxidized osmate with a layered perovskite structure shows fascinating magnetism in combination with strong lattice dynamics.

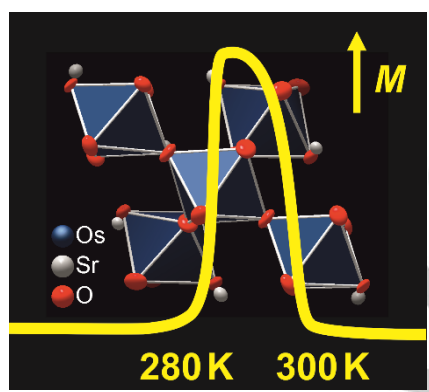

*Prof. Dr. Gohil S. Thakur\*, Dr. Julia-Maria Hübner, Dr. Kati Finzel, Prof. Dr. Thomas Doert, Prof. Dr. Michael Ruck*

*Page No. – Page No.*

**$Sr_4Os_3O_{12}$ — A Layered Osmate(V,VI) that is Magnetic close to Room Temperature**

Additional Author information for the electronic version of the article.

Gohil S. Thakur: 0000-0002-1362-2357
Julia-Maria Hübner: 0000-0003-2048-6629
Kati Finzel: 0000-0002-0862-2782
Thomas Doert: 0000-0001-7523-9313
Michael Ruck: 0000-0002-2391-6025